\documentclass[aps,prl,showpacs,superscriptaddress,twocolumn]{revtex4-2}
\usepackage{amsmath}
\usepackage{amssymb}
\usepackage{xcolor}
\usepackage{graphicx}% Include figure files
\usepackage{dcolumn}% Align table columns on decimal point
\usepackage{bm}
\usepackage{natbib}
\usepackage{float}
\usepackage{ulem}

\newcommand{\ket}[1]{\mbox{$ | #1 \rangle$}}

\begin{document}
	
%	\preprint{APS/123-QED}
	
	\title{$D_1$ Magic Wavelength Tweezers For Scaling Atom Arrays}
	
	\author{Mohammad Mujahid Aliyu}
	\affiliation{Centre for Quantum Technologies, National University of Singapore, 117543 Singapore, Singapore}
	\author{Luheng Zhao}
	\affiliation{Centre for Quantum Technologies, National University of Singapore, 117543 Singapore, Singapore}
	\author{Xiu Quan Quek}
	\affiliation{Centre for Quantum Technologies, National University of Singapore, 117543 Singapore, Singapore}
	\author{Krishna Chaitanya Yellapragada}
	\affiliation{Centre for Quantum Technologies, National University of Singapore, 117543 Singapore, Singapore}
	\author{Huanqian Loh}
	\email[]{phylohh@nus.edu.sg}
	\affiliation{Centre for Quantum Technologies, National University of Singapore, 117543 Singapore, Singapore}
	\affiliation{Department of Physics, National University of Singapore, 117542 Singapore, Singapore}
	
%	\date{\today}
	
	\begin{abstract}
		
$D_1$ magic wavelengths have been predicted for the alkali atoms but are not yet observed to date. We experimentally confirm a $D_1$ magic wavelength that is predicted to lie at 615.87~nm for $^{23}$Na, which we then use to trap and image individual atoms with 80.0(6)\% efficiency and without having to modulate the trapping and imaging light intensities. We further demonstrate that the mean loading efficiency remains as high as 74.2(7)\% for a 1D array of eight atoms. Leveraging on the absence of trap intensity modulation and lower trap depths afforded by the $D_1$ light, we achieve an order-of-magnitude reduction on the tweezer laser power requirements and a corresponding increase in the scalability of atom arrays. The methods reported here are applicable to all the alkalis, including those that are attractive candidates for dipolar molecule assembly, Rydberg dressing, or are fermionic in nature.

	\end{abstract}
	
	\maketitle

Optical tweezer arrays with individually trapped atoms have recently emerged as a promising platform for studies of quantum many-body physics, quantum information processing, and metrology \cite{browaeys2020many, morgado2021quantum, graham2019rydberg, madjarov2019atomic, young2020half}. These atom arrays offer the key advantages of rapid loading times, ground-state cooling \cite{kaufman2012cooling, thompson2013coherence}, and generation of defect-free arrays in arbitrary geometries \cite{barredo2016atom, endres2016atom, kim2016situ, ohldemello2019defect, sheng2021efficient, ebadi2021quantum, scholl2021quantum}. When combined with optical lattices, tweezer arrays may be further used to prepare low-entropy states \cite{kaufman2014two, bakr2009quantum, weitenberg2011single, kumar2018sorting}.

To date, most of the literature on tweezer-trapped single atoms has focused on the alkalis Rb and Cs \cite{browaeys2020many, morgado2021quantum, schlosser2001sub, mckeever2003state}, with recent work reported on alkaline earths \cite{madjarov2019atomic, young2020half, saskin2019narrow}. Realizing atom arrays of light alkalis (Li, Na, K) would pave the way for ultracold dipolar molecule studies \cite{carr2009cold} with single-molecule resolution \cite{liu2018building, anderegg2019optical, he2020coherently, cairncross2021assembly, brooks2021preparation}, where molecules that are assembled from pre-cooled atoms typically require a heavy atomic species and a light atomic species to access large electric dipoles. Further, some of these alkalis have small hyperfine splittings that allow them to be doubly Rydberg-dressed with simpler laser setups \cite{guardadosanchez2021quench, lorenz2021rydberg}, making them attractive candidates for studies of frustrated quantum magnets \cite{glaetzle2015designing} and symmetry-protected topological phases \cite{potirniche2017floquet}. Tweezer arrays of fermionic alkalis with tunable Feshbach resonances can also be used to explore assembled systems of Fermi-Hubbard physics \cite{murmann2015two, esslinger2010fermi} in flexible geometries of relevance to condensed matter physics.

Yet, despite the promising applications of light alkalis, efforts to trap single atoms directly from molasses have been impeded by high heating rates from their low mass and/or anti-trapping excited $P_{3/2}$ state. To circumvent these problems, one can use ``AC tweezers'', where the trapping and $D_2$ cooling light are modulated out of phase at a frequency much faster than the tweezer trap frequency. This technique was found to be necessary in Na \cite{hutzler2017eliminating} and K \cite{lorenz2021raman} to achieve single-atom trapping. We note that in Rb, AC tweezers have also been recently employed to image single atoms without problematic light shifts \cite{brown2019gray}. However, AC tweezers come at the cost of greater trap laser power requirements. Firstly, to achieve a given trap depth, AC tweezers require power that is larger by a factor of the inverse of the duty cycle (relative to an unmodulated trap). Secondly, AC tweezers on the $D_2$ line typically require deeper traps to load and image single atoms. These power constraints impede the scalability of atom arrays.

On the other hand, magic wavelengths that cancel the differential light shift of the alkali $D_1$ transition have been predicted \cite{arora2007magic, UDportal} but never observed. Furthermore, these wavelengths tend to yield relatively large and positive polarizabilities, which facilitates easy trapping of alkalis. Such user-friendly $D_1$ magic wavelengths exist for all the alkalis, unlike the case for $D_2$ magic wavelengths that only exist on the closed cycling transition for Cs \cite{mckeever2003state, phoontong2010characterization, liu2017measurement, yoon2019laser}. Additionally, the $D_1$ transition has been used for efficient cooling and high fidelity imaging \cite{fernandes2012sub, grier2013lambda, colzi2016sub, cheuk2018lambda} using $D_1$ $\Lambda$-enhanced gray molasses ($\Lambda$GM). The $D_1$ light in the tweezer-loading stage has also been used to enhance the single-atom loading probability \cite{grunzweig2010near, carpentier2013preparation, fung2015efficient, lester2015rapid} in shallow trap depths \cite{brown2019gray} by carefully tailoring atom collisions. These efficient loading and imaging techniques eliminate the need for AC tweezers and their challenging power demands.

\begin{figure}[tb]
\includegraphics[width=8.2cm]{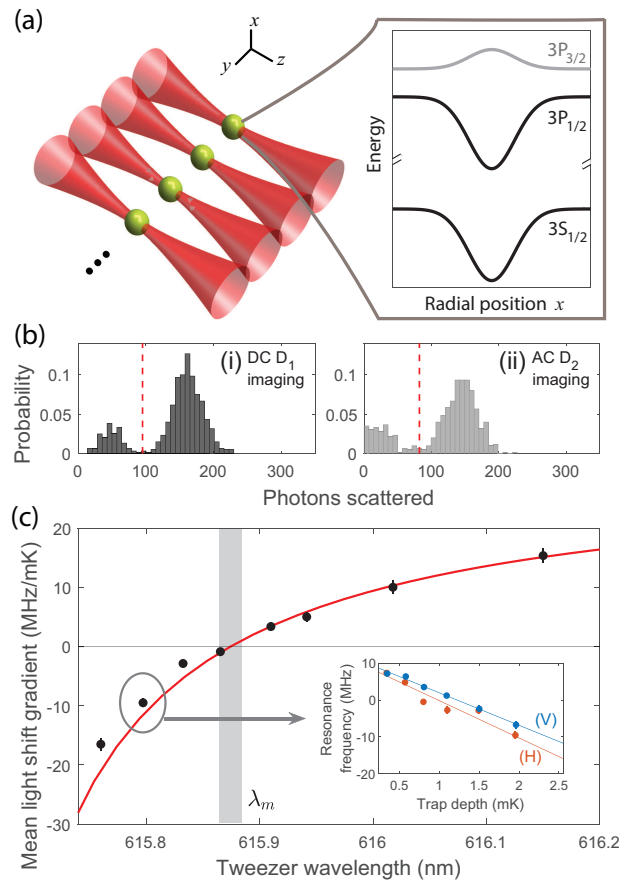}
\caption{\label{figure1}(a) A 1D array of trapped single $^{23}$Na atoms. At the $D_1$ magic wavelength 615.87~nm, the ground $3 ^{2}S_{1/2}$ and first excited $3 ^{2}P_{1/2}$ states experience the same light shifts whereas the $3 ^{2}P_{3/2}$ state is slightly anti-trapping. (b) Histograms of photons scattered using (i) DC $\Lambda$-enhanced $D_1$ imaging and (ii) AC $D_2$ imaging. Photons scattered to the right (left) of the red dashed line indicate one (no) atom present in the trap. (c) Experimental confirmation of the theoretically predicted magic wavelength $\lambda_m$ (vertical gray line) for the $D_1$ transition. The width of the gray line reflects the error bar from \cite{arora2007magic}. The theory curve without free parameters (red) is overlapped with the experimental data points (black), which shows the $D_1$ light shift normalized by the trap depth as a function of tweezer wavelength. The inset shows data taken at a tweezer wavelength of 615.80~nm for 6 different trap depths at two linear orthogonal input polarizations of the trap (horizontal $H$ and vertical $V$). The light shift gradients of the two polarizations are averaged to obtain the data point circled in the main figure.}
\end{figure}

We report here an experimental confirmation of a $D_1$ ($3 ^{2}S_{1/2} \rightarrow 3 ^{2}P_{1/2}$) magic wavelength for $^{23}$Na that is predicted to lie at 615.87~nm, which we use to achieve high loading probabilities for single atoms (80.0(6)\%) and for a 1D array of eight atoms (74.2(7)\%) [Fig.~\ref{figure1}a]. Operating the tweezer at the magic wavelength yields the highest $D_1$ imaging efficiency. This allows us to develop an all-DC scheme to load and image single atoms using $D_1$ transitions, as opposed to using the $D_2$ line where AC tweezers are observed to remain necessary for single-atom loading. Compared to AC $D_2$ imaging, DC $D_1$ imaging can be efficiently carried out in traps as shallow as 0.5~mK, which requires about ten times less tweezer power. By using the all-DC approach, we generate an order of magnitude more traps for a given tweezer power, which leads to a corresponding increase in atom-array scalability. 

We begin our experiments by loading a magneto-optical trap, performing polarization gradient cooling with $D_2$ light, and then applying $\Lambda$GM for 6.8~ms on the $D_1$ transition to further cool the atoms to 15(1)~$\mu$K. During the $\Lambda$GM stage, we also turn on the tweezer trap formed from focusing a 616~nm laser through a high numerical aperture microscope objective (NA = 0.67) to a waist of 0.73~$\mu$m. Both the $D_1$ and tweezer light are kept on for another 100~ms, during which the $D_1$ light is blue-detuned to roughly 1.5 times the tweezer trap depth, to activate an enhanced loading mechanism \cite{brown2019gray, grunzweig2010near, carpentier2013preparation, fung2015efficient, lester2015rapid}. Finally, we detect the atoms using $\Lambda$-enhanced imaging on the $D_1$ line, where the imaging light is also blue-detuned from the single-photon resonance. All of the above steps are done without any modulation of the tweezer light \cite{supplinfo}.

If the loading process is not complete after 100~ms, multiple atoms may be left in the trap, all of which can be detected using $\Lambda$-enhanced $D_1$ imaging light, which does not modify the trap occupation \cite{mcgovern2011counting}. In contrast, using red-detuned imaging light ensures the occupation of at most one atom in the trap after imaging due to light-assisted collisions in the collisional-blockade regime. Therefore, to check if the loading process is complete, we compare the histogram taken with blue-detuned $\Lambda$-$D_{1}$ imaging light [Fig.~\ref{figure1}b(i)] against that taken with red-detuned $D_2$ imaging light that is modulated out-of-phase with the tweezer light at an AC frequency of 2.5~MHz [Fig.~\ref{figure1}b(ii)]. Both histograms are similar and show only two peaks. The red dashed line indicates the photon threshold above (below) which one (no) atom is detected. This allows us to confirm that the loading process is complete and we are only imaging single atoms with the $\Lambda$-enhanced $D_1$ light.

We first experimentally confirm the predicted $D_1$ magic wavelength by probing the differential light shift experienced by an atom in a single optical tweezer. The light-shift probe is a linearly-polarized beam that propagates along the -$x$+$z$ direction with total intensity 8.2~$I_{\text{sat}}$ ($I_{\text{sat}}$ = 6.26~mW/cm$^{2}$). Its principal detuning from the $3 ^{2}S_{1/2} \ket{F = 2} \rightarrow 3 ^{2}P_{1/2} \ket{F' = 2}$ transition $\Delta_{2 \rightarrow 2'}$ is scanned, such that when on resonance with the light shifted transition, the probe beam heats the atom out of its tweezer trap in a few hundred microseconds. To increase the scattering rate and avoid hyperfine dark states, the probe beam contains repumper light that is off-resonant from the $3 ^{2}S_{1/2} \ket{F = 1} \rightarrow 3 ^{2}P_{1/2} \ket{F' = 2}$ transition by a Raman detuning of $\delta = \Delta_{1 \rightarrow 2'} - \Delta_{2 \rightarrow 2'} = -20$~MHz, where $I_{\text{repump}}/I_{\text{cool}} = 0.07$ \cite{supplinfo}. The atom also experiences a magnetic field of 1.2~G, which helps lift the degeneracy between $m_F$ states. To maintain efficient imaging regardless of the tweezer wavelength, we use AC $D_2$ imaging to detect the single atom after the probe beam is turned off. 

Each probe scan result is fit to a Lorentzian to obtain the resonance frequency. To reduce the effect of magnetic field fluctuations contributing to a resonance frequency shift, we repeat the scan at 5-6 trap depths for a given tweezer wavelength and determine the gradient of the resonance frequency versus trap depth. We further suppress the vector polarizability contribution from residual circular polarization of the trap light by averaging the light-shift gradient data taken with two orthogonal linear polarizations of the tweezer light [Fig.~\ref{figure1}c, inset] \cite{herold2012precision}. Since tensor polarizability components are absent for the ground and excited $J = 1/2$ states, the averaged light-shift gradient is dominated by the scalar light shift of the $D_1$ transition. 

Figure~\ref{figure1}c summarizes the light-shift gradients measured at different tweezer wavelengths. The experimental data agrees well with the theory curve obtained with no free parameters \cite{arora2007magic, UDportal}. The predicted magic wavelength $\lambda_m$, at which the differential light shift is zero, is confirmed to be 615.87~nm. Over the range of tweezer wavelengths probed, theory predicts that the $3 ^{2}S_{1/2}$ ground state light shift remains the same to within 2\%. Therefore, the decrease (increase) in differential light shift at wavelengths $\lambda < \lambda_{m}$ ($\lambda > \lambda_{m}$) is mainly attributed to atoms experiencing a deeper (shallower) trap when in the $3 ^{2}P_{1/2}$ excited state. 

\begin{figure}[tbp]
  \centering
\includegraphics[width=8.2cm]{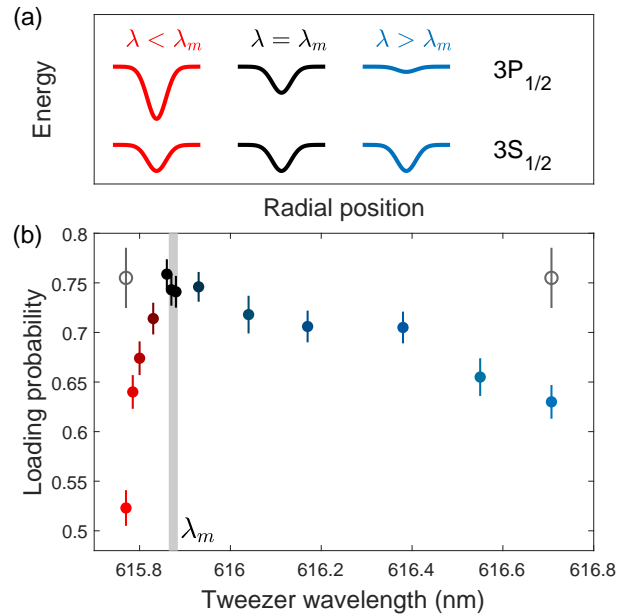}
\caption{(a) Differential light shifts of the $3 ^{2}S_{1/2}$ and $3 ^{2}P_{1/2}$ states on either side of the magic wavelength $\lambda_m$. (b) Single-atom loading probabilities, achieved with $D_1$ trap loading and either AC $D_2$ imaging (open circles) or DC $D_1$ imaging (solid circles), as a function of tweezer wavelength close to $\lambda_m$ (vertical gray line). The reported loading probability includes both the loading efficiency and imaging efficiency.}
\label{figure2}
\end{figure}

Away from the magic wavelength, the change in differential light shift at different radial positions of the trap [Fig.~\ref{figure2}a] can potentially affect the DC trap-loading and imaging of single atoms, both of which use the $D_1$ transition. We first investigate the effect on trap loading by optimizing the loading detunings for a 1.0~mK trap and subsequently detecting the atoms with AC $D_2$ imaging in a deeper trap. We find that, as long as we compensate for the change in differential light shift at the trap center, the loading efficiency remains unaffected by the tweezer wavelength, which implies that the loading process cools the atom to the trap center [open gray circles in Fig.~\ref{figure2}b].  

On the other hand, the tweezer wavelength has a strong effect on the $D_1$ imaging efficiency, which is maximized only at the magic wavelength [solid circles in Fig.~\ref{figure2}b]. $D_1$ imaging is implemented using a pair of counter-propagating laser beams (lin-perp-lin polarized, total intensity 59 $I_{\text{sat}}$, $I_{\text{repump}}/I_{\text{cool}}$ = 0.07). At each tweezer wavelength, the $D_1$ imaging detuning is scanned to optimize the efficiency. The sensitivity of the $D_1$ imaging efficiency to the tweezer wavelength indicates that the atom heats up during imaging, such that it samples different radial positions of the trap, giving rise to inhomogeneous detunings that are difficult to compensate for. We note that there is a region of $\sim$0.05~nm around the magic wavelength that retains the high imaging efficiencies in an imaging trap depth of 1.4~mK. This wavelength region is expected to become broader if the imaging trap depth is allowed to decrease.

\begin{figure*}[htbp]
  \centering
\includegraphics[width=16.5cm]{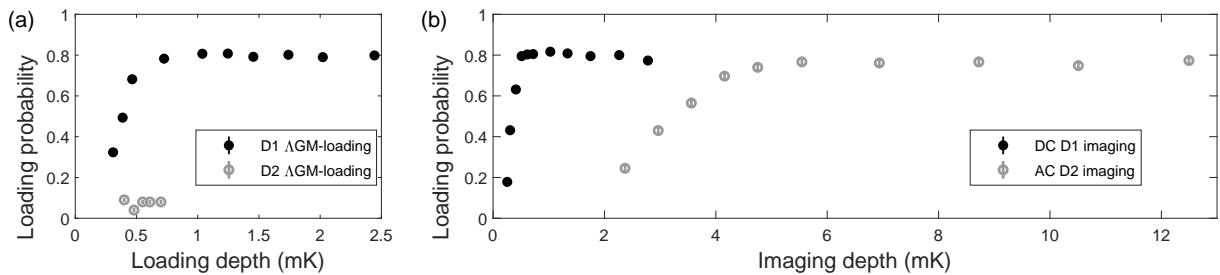}
\caption{(a) Enhanced loading probabilities obtained with $D_1$ $\Lambda$GM-loading (solid circles) and suppressed loading probabilities obtained with $D_2$ $\Lambda$GM-loading (open circles). (b) Imaging trap depth comparison for DC $D_1$ imaging (solid circles) versus AC $D_2$ imaging (open circles). For AC imaging, the $x$-axis refers to the peak value of the trap depth.}
\label{figure3}
\end{figure*}

To take full advantage of our magic wavelength results, we determine the lowest loading trap depth and imaging trap depth at which we can operate to maintain a high loading probability. With the tweezer wavelength fixed at 615.87~nm, we optimize the $D_1$ detunings and intensities, first for different trap depths at which the 100~ms loading stage occurs, and subsequently for the trap depths at which $\Lambda$-enhanced $D_1$ imaging occurs. For the former, the loading probability is optimized when the $D_1$ light is blue-detuned by 1.5-1.6 times the trap depth \cite{fung2015efficient, brown2019gray}. At this detuning, the $D_1$ light is able to cool the atoms while driving them to repulsive molecular states, leading to an enhanced single-atom loading efficiency that saturates at 80\% when the trap depth is above 0.7~mK [solid circles in Fig.~\ref{figure3}a].

For comparison, we replace $D_1$ $\Lambda$GM-cooling and loading of the optical tweezers with $D_2$ $\Lambda$GM light \cite{shi2018sub, rosi2018lambda, hsiao2018lambda}, still in DC mode. For $D_2$ $\Lambda$GM cooling to work, the $D_2$ light needs to be nominally blue-detuned from the $3 ^{2}S_{1/2} \ket{F = 2} \rightarrow 3 ^{2}P_{3/2} \ket{F'' = 2}$ transition, yet remain somewhat far-detuned from the $3 ^{2}S_{1/2} \ket{F = 2} \rightarrow 3 ^{2}P_{3/2} \ket{F'' = 3}$ that lies only 58~MHz above the former transition in free space. This allows us to attain an atom temperature of 50(2)~$\mu$K with $D_2$ $\Lambda$GM cooling without tweezer light present \cite{supplinfo}. Yet, we are unable to load a single atom into the optical tweezer at 615.87~nm more than 10\% of the time [open circles in Fig.~\ref{figure3}a], despite varying the detunings, intensities and durations over a wide range. We attribute this failure to load to the breakdown of the $D_2$ hyperfine structure in the presence of the tweezer light, complicated by the tensor polarizability of the excited $3 ^{2}P_{3/2}$ state \cite{lekien2013dynamical, neuzner2015breakdown, supplinfo}. The mixing of these Zeeman-hyperfine states inhibits efficient loading of a single atom into the tweezer trap. Therefore, we confirm that AC tweezers are still required with the $D_2$ line. 

We also compare the trap depths at which the imaging efficiencies saturate for DC $D_1$ imaging versus AC $D_2$ imaging. In both cases, the trap depth during the loading stage is kept constant at 0.92~mK before ramping to the desired trap depth for imaging. The number of photons scattered per atom is kept the same for both types of imaging. For DC $D_1$ imaging, the probability of loading and detecting a single atom saturates at (80.0 $\pm$ 0.6)\% as long as the imaging trap depth is above 0.5~mK [solid circles in Fig.~\ref{figure3}b], whereas the peak trap depth required for saturation with AC $D_2$ imaging is about ten times higher [open circles in Fig.~\ref{figure3}b]. For AC imaging, the peak imaging trap depth takes into account the AC tweezer duty cycle of 30\%, which is chosen to optimize detection efficiency and is consistent with past efforts to load single $^{23}$Na atoms in AC tweezers \cite{hutzler2017eliminating, supplinfo}. By reporting the peak imaging trap depth instead of the time-averaged imaging trap depth, we emphasize the fact that a given amount of tweezer laser power creates a higher number of tweezer traps with an all-DC $\Lambda$-enhanced $D_1$ approach compared to using AC tweezers with $D_2$ light.

\begin{figure}[htbp]
  \centering
\includegraphics[width=8.2cm]{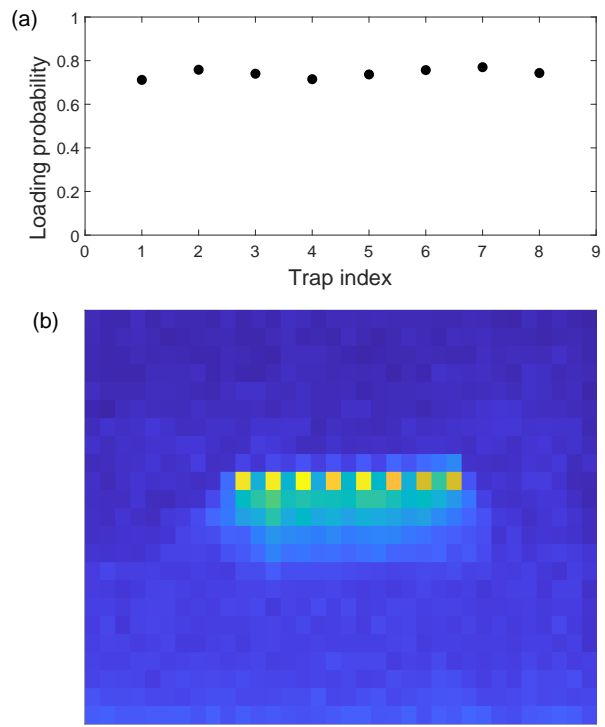}
\caption{(a) Enhanced loading probabilities for a 1D array of eight $^{23}$Na atoms. (b) Fluorescence image of the 1D atom array.}
\label{figure4}
\end{figure}

To illustrate the applicability of this work, we demonstrate that the enhanced loading probabilities extend to a one-dimensional array of eight atoms separated by 2.4~$\mu$m. Each tweezer trap of 1.1(1)~mK depth is generated by sending the tweezer laser through an acousto-optic deflector, such that 2.4~mW per trap is incident on the atoms and focused to a waist of 0.78~$\mu$m. Using the same loading and imaging parameters as that optimized for a single tweezer, we achieve a loading efficiency of (74.2 $\pm$ 0.7)\% [Fig.~\ref{figure4}a].	

In conclusion, using the magic wavelength for the $D_1$ transition, we have developed an all-DC scheme to load and image single atoms at an enhanced probability of 80\% and at reduced trap depths. These methods allow us to scale up the number of tweezer traps for a given laser power by an order of magnitude. We note that while the minimum loading trap depth at which the loading probability saturates is 0.7~mK, the imaging trap depth for efficient detection can be as low as 0.5~mK. Since both loading and imaging processes make use of $\Lambda$-$D_1$ transitions, we expect that the loading trap depth can be further optimized to lower values with a detailed understanding of the loading dynamics, thereby further enhancing the scalability of atom arrays. 

We also note that throughout this work, the reported loading efficiencies are significantly higher than the loading efficiency of 50-60\% typically obtained with red-detuned $D_2$ molasses \cite{schlosser2002collisional}. Nevertheless, our highest observed loading probability of 82(2)\% lies below the state-of-the-art enhanced loading probability of 89(1)\% achieved with $D_1$ gray-molasses in shallow traps for Rb \cite{brown2019gray}. This discrepancy is most likely due to our decision to minimize the atom temperature prior to the 100~ms loading stage. The low initial temperature in turn suppressed the uneven distribution of kinetic energy between the two atoms in a collision pair, which is a precursor for activating the mechanism where the more energetic atom is kicked out and leaves one remaining atom in the trap.

The magic wavelength-enabled methods demonstrated here in DC are realized with an atomic species ($^{23}$Na) that previously required the modulation of tweezer light \cite{hutzler2017eliminating} in order to achieve single-atom trapping. Given that user-friendly $D_1$ magic wavelengths exist for all the alkali atoms \cite{arora2007magic, UDportal} and that AC tweezers are so far required to trap other light alkalis like potassium \cite{lorenz2021raman}, we expect these techniques to benefit in particular the light alkalis. The pursuit of atom arrays of these species would allow us to access rich interactions through dipolar molecule assembly \cite{carr2009cold}, Rydberg dressing \cite{browaeys2020many}, or Feshbach resonances \cite{chin2010feshbach}. Beyond tweezer arrays, these methods also extend to optical lattice systems \cite{nelson2007imaging, bakr2009quantum, bloch2012quantum, heinz2020state}. The ability to establish single-particle control on a wide variety of atoms and molecules in a scalable manner is a key step for applications in quantum information processing, quantum many-body physics studies, and precision measurement.

\begin{acknowledgments}
We thank the group of Kang-Kuen Ni for helpful discussions on the AC trapping of single $^{23}$Na atoms, and Swarup Das, Jun Ming Billy Lim, Tianli Lee, and Weihong Yeo for experimental assistance. This research is supported by the National Research Foundation (NRF) Singapore, under grant no.\ NRF-NRFF2018-02, and the Ministry of Education, Singapore.
\end{acknowledgments}

%%%%%%%%%%%%%%%%%%%%%%%%%%%%%%%%%%%%%%%%%%%%%%%%%%%%%%%%%%%%%%%%%%%%%%%%%%%%
%%\bibliographystyle{apsrev4-2}
%%	\bibliography{references}

%

%%%%%%%%%%%%%%%%%%%%%%%%%%%%%%%%%%%%%%%%%%%%%%%%%%%%%%%%%%%%%%%%%%%%%%%%%%%%%

\end{document}